\documentclass[12pt]{article}

\usepackage{enumerate}
\usepackage{amssymb,amsmath,amsfonts,amsthm,textcomp}
\usepackage{eucal,graphics}
\usepackage{endnotes}
\usepackage{natbib} 
\usepackage{setspace}

\newcommand{\e}{\ensuremath{\mathbb{E}}}

\newcommand{\p}{\ensuremath{\mathbb{P}}}
\newcommand{\q}{\ensuremath{\mathbb{Q}}}

\newcommand{\spa} {\ensuremath{\mbox{ }}}

\newcommand\ent[2]{ \citep[#2]{#1}}

\begin{document}

\title{Ethics and Finance: the role of mathematics\footnote{Research supported by
EPSRC grant \, EP/C508882/1}\ 
%\footnote{This is an Author's Original Manuscript of an article submitted for consideration in the Review of Social Economy  [copyright Taylor \& Francis].}
}

\author{
{\sc Timothy C.\,Johnson\footnote{Maxwell Institute for
Mathematical Sciences and Department of Actuarial Mathematics
and Statistics, Heriot-Watt University, Edinburgh EH14 4AS,
UK, \texttt{t.c.johnson@hw.ac.uk}}}}

\maketitle

%\doublespacing
\begin{abstract}
This paper presents the contemporary Fundamental Theorem  of Asset Pricing as being equivalent to approaches to pricing that emerged before 1700 in the context of Virtue Ethics.  This is done by considering the history of science and mathematics in the thirteenth and seventeenth century.  An explanation as to why these approaches to pricing were forgotten between 1700 and 2000 is given, along with some of the implications on economics of viewing the Fundamental Theorem as a product of Virtue Ethics.

The Fundamental Theorem was developed in mathematics  to establish a `theory' that underpinned the Black-Scholes-Merton approach to pricing derivatives.  In doing this, the Fundamental Theorem unified a number of different approaches in financial economics, this strengthened the status of neo-classical economics based on Consequentialist Ethics.  We present an alternative to this narrative.
\end{abstract}
\section{Introduction}

The final conclusion of The {Financial} {Crisis} {Inquiry} {Report} was that in the lead up to the financial crisis of 2007--2009 (hereafter `FC') there had been a  ``systemic breakdown in accountability and ethics'' %and that ``to pin this crisis on mortal flaws like greed and hubris would be simplistic. It was the failure to account for human weakness that is relevant to this crisis.'' 
\citep{FCIC}.  This article explores this breakdown in ethics  from a perspective based in  mathematics. Specifically, we argue that the root of the problem lay in Consequentialist Ethics developed in response to scarcity in a broadly deterministic environment, when what was required were practices rooted in Virtue Ethics developed in response to uncertainty.  

This argument is made by presenting the  Fundamental Theorem of Asset Pricing (hereafter `FTAP') as having its foundations in the concept of balanced reciprocity \ent{S_SAE}{pp 193--196} and Virtue Ethics, which we argue are  responses to uncertainty, rather than in neo-classical axioms of market prices and Consequentialist Ethics.  This case is made by comparing the essential elements of the FTAP with Scholastic discussions of the `just price' in the thirteenth century and then the development of probability theory in the context of the fair pricing of commercial contracts in the second half of the seventeenth century.  We identify these two periods as being dominated by uncertainty, where as the emergence of Consequentialist Ethics took place in an environment of scarcity.  We finish the article by giving some of the impacts on financial economics of approaching the FTAP from the perspective of Virtue Ethics.

The main contribution of this paper is in providing an alternative narrative around the FTAP to compete with the  narrative that it unifies a number of distinct themes that emerged in financial economics in the 1960s: the Efficient Markets Hypothesis; the Capital Asset Pricing Model;  the Arrow–Debreu model; and the use of stochastic processes to model a dynamic economy.  The unification of these by the   FTAP enhances their status, giving them the stature of the `immutability and indubitability of mathematics' and supports the  paradigm of value-neutral economics.  Presenting a different narrative undermines this and the neo-classical framework in general.

The argument presented here places Virtue Ethics at centre of financial economics.  In doing this the paper is aligned to modern work on Virtue Ethics originating in Anscombe \citep{A_MMP} and developed by, amongst others, McCloskey \citep{M_BV} and van Staveren \citep{vS_BUD}. It also supports Putman's advocacy of economics without dichotomies \citep{P_EEWD}, by considering the `fact' of a price within an  ethical framework.  Moreover, we suggest that the arguments presented go some way to address the `problem' of embeddeness in economics discussed by Granovetter  \citep[p 490]{G_EASS}.

The case is made that Virtue Ethics, centred on an agent's character rather than the act or its consequences, is desirable when dealing with epistemic uncertainty and essential in the face of ontological uncertainty. This creates links with Post-Keynesian theories \citep{D_RET}, specifically we add an ethical interpretation to probability omitted in Davidson's separation of modern probability  into subjectivist or objective theories \citep{D_PTRU}.  Furthermore, the emphasis on reciprocity and hence the centrality of the credit-debit relations imply the non-neutrality of money advocated by Post Keynesians \citep{I_MSR}.  

Finally, central to the whole argument is the role economics has had in generating mathematics and the subsequent mathematisation of science, and supports the arguments of The Dahlem Group on Economic Modeling.   In effect, in presenting an alternative to the neo-classical narrative behind the FTAP, the paper links different areas of heterodox economics.

While the motivation of the article is in economics, the argument is based substantially on comparing  modern and historical  mathematical constructs.  This runs the risk presenting a Whig history; that mathematicians in the past  thought as modern mathematicians do, demonstrating the immutability, and hence, the indubitability, of mathematics.  However, the objective of this paper is to demonstrate the reverse.  Today, the general understanding of the FTAP is very different from how a seventeenth century mathematician would have interpreted it.  The explanation for this difference is given as a change in social environment and this means that this article offers an example of the `social construction' of mathematics (e.g. \citep{R_MSSP}, \citep{H_WMR}, \citep{G_SCMC} \citep{RB_WTM})

\section{The Fundamental Theorem of Asset Pricing}

Within the field of Financial Mathematics, the Fundamental Theorem of Asset Pricing consists of two statements, (e.g. \citep[Section 5.4]{S_SCF2}) \\
\textsc{Theorem: The Fundamental Theorem of Asset Pricing}
\begin{enumerate}
 \item  \textit{A market admits no arbitrage, if and only if, the market has a martingale measure.}
 \item \textit{Every contingent claim can be hedged, if and only if, the martingale measure is unique.} 
\end{enumerate}

The theorem emerged between 1979 and 1983 (\citep{HK_MAMSM}, \citep{HP1_FT},\citep{HP2_FT}) as Michael Harrison sought to establish a mathematical theory underpinning the well established Black-Scholes equation for pricing options.  One remarkable feature of the FTAP is its lack of mathematical notation, which is highlighted by the use of mathematical symbols in the Black-Scholes equation, which came out of economics.  Despite its non-mathematical appearance, the work of Harrison and his collaborators opened finance to investigation by functional analysts (such as \citep{S_UFG}) and by 1990, any mathematician working on asset pricing would have to do so within the context of the FTAP.

The use of the term `measure' in relation to probability places the FTAP within the theory of probability  formulated by Andrei Kolmogorov in 1933 \citep{K_FTP}.  Kolmogorov's work took place in a context captured by Bertrand Russell, who in 1927 observed that 
\begin{quote}
It is important to realise the fundamental position of probability in science. \ldots As to what is meant by probability, opinions differ. {\citep[p 301]{OoP}}
\end{quote}

The significance of probability in providing the basis of statistical inference in empirical science had been generally understood since the 1820s following the two texts on probability published by  Laplace.  In the 1920s the idea of randomness, as distinct from a lack of information, the absence of Laplace's Demon, was becoming significant.  In 1926 the physicist Max Born was ``inclined to give up determinism'', to which Einstein responded with  ``I for one am convinced that [God] does not play dice'' { \citep[pp 147--157]{vP_CMP}}.  Outside the physical sciences, Frank Knight, in \emph{Risk, Uncertainty and Profit},  argued that uncertainty was the only  source of profit, since if a profit was predictable the market would respond and make it disappear \citep[III.VII.1--4]{RUP}.  Simultaneously, in his \emph{Treatise on Probability}, John Maynard Keynes observed that in some cases cardinal probabilities could be deduced, in others, ordinal probabilities, one event was more or less likely than another, could be inferred, but the largest  class of problems  in economics were not reducible to the mathematical concept of probability \citep[ Ch XXIV, 1]{K_TP}.  The distinction between Knight's and Keynes' approach is that Knight sees uncertainty as a problem of epistemology, we cannot know everything, where as Keynes sees the problem as one of ontology, numerical probability of economic events in the future might not exist.

Two mathematical theories of probability had become ascendant by the late 1920s. Richard von Mises, an Austrian engineer linked to the Vienna Circle of logical-positivists, and brother of the economist Ludwig, attempted to lay down the axioms of probability based on   observable facts  within a framework of Platonic-Realism.  The result was  published in German in 1931  and popularised in English as \emph{Probability, Statistics and Truth} and is  now regarded as a key justification of the frequentist approach to probability.  

To balance von Mises' Realism, the Italian actuary, Bruno de Finetti presented a more Nominalist approach.  De Finetti  argued that ``Probability does not exist'' because it was only an expression of the observer's view of the world.  De Finetti's subjectivist approach was closely related to the less well-known position taken by Frank Ramsey, who, in 1926, published \emph{Probability and Truth}, his argument was well-received by his friend and mentor John Maynard Keynes, but his early death hindered its development outside philosophy.

While von Mises and de Finetti took an empirical path, Kolmogorov constructed his theory on the basis of deduction from axioms, the classical `Greek' approach to mathematics \citep[pp 44--45]{F_CPL}.  Kolmogorov wanted to address two issues: the concept that probability changes and that there needed to be a theory that accommodated infinite possibilities.  The thesis of Louis Bachelier is well known following its promotion by Samuelson as a precursor of the Black-Scholes equation,  this status is debatable  when the activity of nineteenth century investors and actuaries is considered  \citep{P_IPRI}. Bachelier's thesis did make a significant contribution in that it introduced the idea that probability distributions evolved in time (\citep[p 46]{B_TS}, \citep[p 40]{TdS}).  This idea was picked up by Kolmogorov, and one contemporary American reviewer noted that Kolmogorov provided an important proof of Bayes' Theorem \citep{R_RGW}, then still controversial \citep[Ch XVI, 13]{K_TP} but now a cornerstone of statistical decision making.  This  is significant in the context of this paper in that it highlights how probability theory has evolved after Knight and Keynes: when Knight classified probability he would have done so within a framework that regarded distributions as static.  Having made this point, we can also note that modern theory still maintains this principle, arguing that decisions change as information changes.  Chris Rogers and Angus Brown, working within Kolmogorov's framework, have argued that it is more convenient to address price changes by modelling changes in `beliefs', captured by probability, rather than `information', carried in the filtration \citep{BR_DB}.

In the second half of the nineteenth century, mathematics began to focus on infinite sets while physics turned its attention to statistical physics, based on large (infinite) ensembles of particles.   Following the work of Montmort  and de Moivre in the first decode of the eighteenth century, mathematical probability had been associated with counting events and comparing relative frequencies, this is incohernet if the set of possibilities is infinite.  Von Mises had tried to address these issues but his analysis was weak.  As Jan von Plato observes
\begin{quote}
von Mises's theory of random sequences has been remembered as something to be criticized: a crank semi-mathematical theory serving as a warning of the state of probability [at the time] {\citep[p 180]{vP_CMP}}
\end{quote}

In 1902 Lebesgue had redefined the mathematical concept of the integral in terms of abstract `measures'  in order to accommodate new classes of mathematical functions that had emerged in the wake of Cantor's transfinite sets.  Kolmogorov made the simple association of these abstract measures with probabilities, solving  von Mises' problem of having to deal with infinite sets in an \emph{ad hoc} manner.  As a result Kolmogorov  identified a random variable with a function and an expectation with an integral: probability became a branch of Analysis, not Statistics.

Kolmogorov's work was initially well received, but slow to be adopted.   Amongst English-speaking mathematicians, the American Joseph Doob was instrumental in promoting probability  as measure \citep{D_PM} while the full adoption of the approach   followed its advocacy by Doob and William Feller at the First Berkeley Symposium on Mathematical Statistics and Probability in 1945--1946. 

While measure theoretic probability is  rigorous, outside pure mathematics it is often seen as  redundant.  Von Mises criticised it as un-necessarily complex \citep[p 99]{PST} while the statistician Maurice Kendall argued that measure theory was fine for mathematicians, but of limited practical use to statisticians and fails ``to found a theory of probability as a branch of scientific method'' \citep[p 102]{K_RTP}.  More recently the physicist Edwin Jaynes champions  Leonard Savage's subjectivism  as having a ``deeper conceptual foundation which allows it to be extended to a wider class of applications, required by current problems of science'' in comparison with measure theory \citep[p 655]{J_PT}.  Furthermore in 2001 two mathematicians Glenn Shafer and Vladimir Vovk, a former student of Kolmogorov, proposed an alternative to measure-theoretic probability, `game-theoretic probability', because the novel approach ``captures the basic intuitions of probability simply and effectively'' \citep{SV_PF}. %Seventy-five years on Russell's enigma appears to be no closer to resolution.

The issue around the `basic intuition' of measure theoretic probability for empirical scientists can be accounted for as a lack of physicality.  Frequentist probability is based on the act of counting, subjectivist probability is based on a flow of information, where as measure theoretic probability is based on an abstract mathematical object unrelated to phenomena.  

Specifically,  in the first statement of the FTAP, the `martingale measure' can be thought of as a probability measure, usually labelled $\q$, such that the price of an asset today, $X_0$, is the expectation, under the martingale measure, of the discounted asset prices in the future, ${X}_T$,
\begin{align*}
 X_0 =\e_\q\big[ {X}_T \big]\ &:= \int_\Omega X_T(\omega)\,\textrm{d}\q(\omega), \quad\omega\textrm{ is an element in a countably infinite set }\Omega,\\
 &:= \sum_{\omega\in\Omega} X_T(\omega)\,\q(\omega), \quad\omega\textrm{ is an element in a finite set }\Omega.
\end{align*}
The probability distribution $\q$ is defined such that an equality exists between the current price, $X_0$, and the expected value of the discounted asset values in the future, $X_T$.  The fact that it is based on current and future prices  means that the martingale measure  is forward looking and the only condition placed on the relationship that the martingale measure has with the `natural', or `physical', probability measure, inferred from historical price changes or subjective judgement, and usually assigned the label $\p$, is that they agree on what is possible.

The term `martingale' in this context derives from doubling strategies in gambling and it was introduced into mathematics by Jean Ville in 1939, in a critique of von Mises work, to label a random process where the value of the process at a specific time is the expected value of the process in the future.  

The concept that asset prices have the martingale property was first proposed by Benoit Mandelbrot \citep{M_FFPUMMM} in response to an early formulation of Eugene Fama's Efficient Market Hypothesis (EMH) \citep{F_BSMP}, the two concepts being combined by Fama in 1970 \citep{F_ECM}.  For Mandelbrot and Fama the key consequence of prices being martingales was that the price today was, statistically, independent of the future price distribution.  In developing the EMH there is no discussion on the nature of the probability under which assets are martingales, and it is often assumed that the expectation is calculated under the natural measure.

%Mathematicians concern themselves with questions of existence and uniqueness.  The existence of a martingale measure is dependent on a lack of arbitrage opportunities in the market, its uniqueness depends on whether or not all claims contingent on asset prices, such as derivatives, can be hedged.

The other technical term in the first statement of the FTAP, arbitrage,  has long been a subject of financial mathematics.  In Chapter 9 of his 1202 text advising merchants, the \emph{Liber Abaci}, Fibonacci discusses `Barter of Merchandise and Similar Things', 
\begin{quote}
 20 arms  of cloth are worth are worth 3 Pisan pounds and 42 rolls of cotton are similarly worth 5 Pisan pounds; it is sought how many rolls of cotton will be had for 50 arms of cloth.\ent{LA}{p 180}
\end{quote}
In this case there are three commodities, arms of cloth, rolls of cotton and Pisan pounds, and Fibonacci solves the problem by having Pisan pounds `arbitrate' between the other two commodities.  

Over the centuries this technique of pricing through arbitration evolved into the law of one price, that if two assets offer identical cash flows then they must have the same price.  This was employed by Jan de Witt in 1671 when he solved the problem of pricing life annuities in terms of redeemable annuities, based on the presumption that
\begin{quote}
the real value of certain expectations or chances of objects, of different value, should be estimated by that which we can obtain from as many expectations or chances dependent on one or several equitable contracts. \citep[p 313, quoting De Witt, ]{S_BECM}%\textit{The Worth of Life Annuities in Proportion to Redeemable Bonds}
\end{quote}
In 1908  Vincent Bronzin published a text  which discusses pricing derivatives by `covering', or hedging them, them with portfolios of options and forward contracts employing the principle of `equivalence',  the law of one price \citep{ZH_VB2}.  In 1965 the functional analyst and probabilist, Edward Thorp, collaborated with a post-doctoral mathematician, Sheen Kassouf, and combined the law of one price with basic techniques of calculus to identify market mis-pricing of warrant prices \citep{M_EW}.  In 1967 they published their methodology in a best-selling book, \emph{Beat the Market}.

Within economics, the law of one price was developed in a series of papers between 1954 and 1964 by Kenneth Arrow,  G\'{e}rard Debreu and Lionel MacKenzie in the context of general equilibrium.  In his 1964 paper, Arrow addressed the  issue of portfolio choice in the presence of risk and introduced the concept of an Arrow Security, an asset that would pay out `1' in a specific  future state of the economy but zero for all other states, and by the law of one price, all commodities could be priced in terms of these securities \citep{A_RSOAR}.  The work of Fischer Black and  Myron Scholes \citep{BS_73} and Robert Merton  \citep{RTORP} employed the principal and presented a mechanism for pricing warrants on the basis that ``it should not be possible to make sure profits'' with the famous Black-Scholes equation being the result.

In the context of the FTAP, `an arbitrage' is the ability to formulate a trading strategy such that the probability, whether under $\p$ or $\q$, of a loss is zero, but the probability of a profit is positive.  This definition is important following Hardie's criticism of the way the term is applied loosely in economic sociology \citep{H_SA}.  The obvious point of this definition is that, unlike Hardie's definition \citep[p 243]{H_SA}, there is no guaranteed (strictly positive) profit, however there is also a subtle technical point: there is no guarantee that there is no loss if there is an infinite set of outcomes.  This is equivalent to the observation that there is no guarantee that an infinite number of monkeys with typewriters will, given enough time,  come up with a work of Shakespeare: it is only that we \emph{expect} them to do so.  This observation explains the caution in the use of infinite sets taken by mathematicians such as Poincar\'{e}, Lebesgue and Brouwer. 

To understand the connection between the financial concept of arbitrage and martingales, mathematical objects, consider  the most basic case of a single period economy,  consisting of a single asset whose price, $X_0$, is known at the start of the period and can take on one of two (present) values, ${X}_T^D<{X}_T^U$, representing two possible states of the economy, at the end of the period. In this case an arbitrage would exist if $X_0\le X_T^D<X_T^U$: buying the asset now, at a  price that is less than the future pay-offs, would lead to a possible profit at the end of the period, with the guarantee of no loss.  Similarly, if ${X}_T^D<{X}_T^U\le X_0$, short selling the asset now, and buying it back at the end of the period would also lead to an arbitrage. %This is the `Dutch book' argument that Ramsey presented and has been discussed widely in the context of philosophy (for example \citep{H_AFAP} and references therein).

 In summary, for there to be no arbitrage opportunities we require that
$$
{X}_T^D\le X_0 \le {X}_T^U.
$$
This implies that there is a real number, $q,\ 0\ge q \ge1$ such that
\begin{align*}
X_0 =\spa& {X}_T^D + q\,({X}_T^U - {X}_T^D)\\
 =\spa& q\,{X}_T^U +(1-q){X}_T^D \\
\equiv\spa& \e_\q\big[ {X}_T \big],
\end{align*}
and it can be seen that $q$ represents a measure theoretic probability that the economy ends in the $U$ state.   If $X_0<X_T^D$ we have that  $q<0$ where as if $X_T^U<X_0$ then $q>1$, and in both cases $q$ does not represent a probability measure, which, by definition must lie between 0 and 1.

With this in mind, the first statement of the FTAP can be  interpreted simply as  ``the price of an asset must lie between its maximum and minimum possible (discounted) future price''. In this simple case there is a trivial intuition behind measure theoretic probability, the martingale measure and an absence of arbitrage are a simple tautology; it is a typical result of Formalist mathematics.

This first statement  was anticipated by Ramsey in 1926.  Ramsey was not satisfied with Keynes' analysis in the \emph{Treatise on Probability} and wished to give a logical argument that probability relations between a premiss and a conclusion exist \citep[p 161]{R_TP}.  He defines `probability' in the subjective sense of `a degree of belief' and makes the point
\begin{quote}
 The old-established way of measuring a person's belief is to propose a bet, and see what are the lowest odds which he will accept. This method I regard as fundamentally sound \citep[p 171]{R_TP}
\end{quote}
He defines some axioms of probability, in particular that a probability lies between 0 and 1 \citep[p 181]{R_TP}.  He goes on to say that
\begin{quote}
These are  the laws of probability, which we have proved to be necessarily true of any consistent set
of degrees of belief. \ldots If anyone's mental condition violated these laws, his choice would depend on the
precise form in which the options were offered him, which would be absurd. He could have a book
made against him by a cunning better and would then stand to lose in any event. \citep[p 182]{R_TP}
\end{quote}
This identifies the absence of the martingale measure with the existence of arbitrage and is today the standard agrument as to why arbitrages do not exist, if they did the, other market participants would bankrupt the agents offering them.  This has become known in philosophy as the `Dutch Book' argument (for example \citep{H_AFAP} and references therein).  Ramsey goes on to say
\begin{quote}
 Having any definite degree of belief implies a certain measure of consistency, namely willingness to
bet on a given proposition at the same odds for any stake, the stakes being measured in
terms of ultimate values. Having degrees of belief obeying the laws of probability implies a further
measure of consistency, namely such a consistency between the odds acceptable on different
propositions as shall prevent a book being made against you. \citep[p 182--183]{R_TP}
\end{quote}
%In this, subjective sense, the martingale measure ensures consistency across the market.  In the context of this paper, it should be noted that Ramsey works within the framework of Consequentialist Ethics, 
%\begin{quote}
%It should be emphasized that in this essay good and bad are never to be understood in any ethical sense but simply as denoting that to which a given person feels desire and aversion.\citep[p 174]{R_TP}
%\end{quote}

To appreciate the meaning of the second statement of the theorem, consider the situation when the economy can take on three states at the end of the time period, not two. If we label possible future  asset prices as ${X}_T^D<{X}_T^M<{X}_T^U $, we cannot deduce a unique set of probabilities $0\le q^U,q^M,q^D \le1$, with $q^U+q^M+q^D=1$, such that
$$
X_0 =q^U\,{X}_T^U +q^M\,{X}_T^M +q^D\,{X}_T^D.
$$
The market still precludes arbitrage, if ${X}_T^D\le X_0 \le {X}_T^U$,  but we no longer have a unique probability measure under which asset prices are martingales, and so we cannot derive unique prices for other assets in the market: Ramsey's consistency is lost.   In the context of the law of one price, we cannot hedge, replicate or cover, a position in the market, making it riskless, while  in terms of Arrow's work the market is incomplete. This explains the sense of the second statement of the FTAP and is important in that it  tells the mathematician that in the real world of imperfect knowledge and transaction costs, a model within the  framework of the FTAP cannot give a precise price.

Most models employed in practice ignore the impact of transaction costs, on the utopian basis that precision will improve as market structures evolve and transaction costs disappear.  Situations where there are many, possibly infinitely many,  prices at the end of the period  are handled by providing a model for asset price dynamics, between times $0$ and $T$.  The choice of asset price dynamics defines the distribution of ${X}_T$, either under the martingale or natural probability measure, and in making the  choice of asset price dynamics,  the derivative price is chosen.  This effect is similar to the choice of utility function determining the results of models in some areas of welfare economics.  A key feature of Black--Scholes--Merton is the choice of log-normal dynamics of asset prices.

The FTAP is not well known outside the limited field of financial mathematics, practitioners focus on the models that are a consequence of the Theorem where as social scientists focus on the original Black-Scholes-Merton model as an exemplar. Practitioners are daily exposed to the imprecision of the models they use and are sceptical, if not dismissive, of the validity of the prices produced by their models  (\citep[pp 409-410 ]{M_BAS}, \citep[p 248]{ENC}, \citep{HT_BS}).  In particular, following the market crash of 1987, few practitioners have used the Black-Scholes equation to actually `price' options, rather they use the equation to measure market volatility, a proxy for uncertainty.

However,  the status of the Black-Scholes model as an exemplar in financial economics has been enhanced following the adoption of  measure theoretic probability. This can be understood because the FTAP, born out of Black-Scholes-Merton, unifies a number of distinct theories in financial economics.  Donald MacKenzie  \citep[p 834]{M_EW} describes a dissonance between Merton's derivation of the model using techniques from stochastic calculus, and Black's, based on the Capital Asset Pricing Model (CAPM).  When measure theoretic probability was introduced it was observed that the Radon-Nikodym derivative, a mathematical  object  that describes the relationship between the stochastic processes Merton  used  in the  natural measure and the martingale measure, involved the market-price of risk (Sharpe ratio), a key object in the CAPM.  This point was well understood in the academic literature in the 1990s and was covered in the fourth edition of the standard text book, Hull's \emph{Options, Futures and other Derivatives}, in 2000.

The realisation that the FTAP unified: Merton's approach, based on stochastic calculus advocated by Samuelson at M.I.T; CAPM, which had been developed at the Harvard Business School and in California;  martingales, a feature of efficient markets that had been proposed at Chicago; and incomplete markets, from Arrow and Debreu in California, enhanced the status of Black-Scholes-Merton as representing a Kuhnian paradigm.   This unification of a plurality of techniques within a  `theory of everything' came just as the Black-Scholes equation was being attacked for not reflecting empirical observations of market prices and obituaries were being written for the broader neoclassical programme (i.e. \citep{C_DNCE}) and goes some way to explain why, in 1997,  the Nobel Prize in Economics was awarded to  Scholes and Merton ``for a new method to determine the value of derivatives''. 

The observation that measure theoretic probability unified a `constellation of beliefs, values, techniques' in financial economics can be explained in terms of the transcendence of mathematics.  To paraphrase Tait 
\begin{quote}
 A mathematical proposition  is about a certain structure, financial markets. It refers to prices and relations among them. If it is true, it is so in virtue of a certain fact about this structure. And this fact may obtain even if we do not or cannot know that it does.  \citep[p 341]{T_TP}
\end{quote}
In this sense, the FTAP confirms the truth of the EMH, or any other of the other `facts' that surround its development.

A less orthodox approach to mathematics is  Nominalist or `humanist': that mathematics is socially constructed.  Within contemporary mathematics this is associated with Reuben Hersh (\citep{H_ME}, \citep{H_WMR}), while in the past it was advocated by Dirk Struik (\emph{On the Sociology of Mathematics}, 1942) and hinted at in Henri Poincar\'{e}'s `occasional pragmatism' \citep{H_HPTPS}, captured in his (in)famous observation 
\begin{quote}
 these two propositions ``the earth turns round,'' and ``it is more convenient to suppose that the earth turns round,'' have one and the same meaning. \citep[p 91]{P_SH}
\end{quote}
The contemporary philosophical case for Nominalism has been made by Charles Chihara \citep{C_SAM}, amongst others.

The aim of the the next two sections of this paper is to develop an understanding of the origins of the FTAP  in the spirit of Nominalist mathematics. In particular, the relationship of the martingale measure and the no arbitrage condition, and the simple tautology described above, shall be explored. The ultimate objective is to provide a theory of asset prices that accommodates ethical principals.

\section{Ethics, Economics and the concept of a `Just Price'}

The conventional taxonomy of ethics identifies three approaches to morality: Deontological, Consequentialist and Virtuous. Deontology can be typified as ``Thou shalt / shalt not'' and guides \emph{action} on the basis of laws, rule or principles. An individual cannot be subject to a law unless it has been promulgated, and so deontology is linked to with philosophical systems that are based on `divine' or ` natural' law, such as Platonism and Stoicism \citep[p 14]{A_MMP}.  The practical problem with Deontological Ethics is that basic rules such as ``Thou shalt not kill'' have caveats while other prohibitions become redundant, or need revising, as society evolves.  In the context of contemporary economics, Deontological Ethics has been employed to ensure market freedom while preventing power--seeking behaviours and in financial regulation (Pillars I \& II of Basel II)  and has been  criticised for being over-bureaucratic and rigid \citep[pp 23--26]{vS_BUD} while susceptible to `gaming' \citep{W_JPMCRM}.

Consequentialism  attempts to judge the value of an action in terms of its \emph{consequences}.  This approach has its roots in ancient Chinese Mohism and Epicureanism, developed in opposition to Platonism and Stoicism.  The approach became fully developed in the nineteenth century  with Bentham, Mill and Sedgewick, who, building on empiricism, argued that one should ``Act always in such a way as to promote the greatest happiness to the greatest number''.  

%These ideas emerged in the eighteenth century in the context in  the Standard Argument Against Free Will: if all acts are pre-determined, an agent cannot be held responsible; if events are random, the agent is also absolved of responsibility for the outcomes of their actions.  The solution to these arguments was to make moral responsibility a premise, rather than the conclusion, of the logical argument.  During this time, it became axiomatic that in order for humans to take  moral responsibility they needed foresight, which could only be obtained through a combination of reason and empiricism: science. If deontology is associated with mathematical Platonism, consequentialism can be associated with mathematical Formalism: ``it is true because we have proved it to be true''.

On the basis of consequentialism and Hume's distinction of `what is' and `what ought to be',   `value--neutrality' was established in economics: since we  have `objective access' to the empirical world' and are `rational beings', we are able to calculate the consequences of our economic actions \citep{WH_EVET}.  One consequence of this faith in calculability was the transformation of economics into a `cyborg science' \citep{M_MDEAC}, and this was not just a feature of neo-classical economics but of many of the research programmes in heterodox economics considered by John Davis (\citep[p 357]{D_TERO}, noting that some biologists view evolution as a deterministic, teleological, process, \citep{TSDE}).

A problem with value-neutrality, described by Robert Heilbroner, is that it misses the critical fact that 
\begin{quote}
 the objects observed by the social scientist all possess an attribute that is lacking in the objects of natural universe.  This is the attribute of consciousness -- of cognition, of ``calculation'', of volition \citep[p 133]{H_EVFS}
\end{quote}
The importance of `volition' had been  recognised by   Oskar Morgenstern, who  objected to perfect foresight based on calculation because
\begin{quote}
always there is exhibited an endless chain of reciprocally conjectural reactions and counter-reactions.  This chain can never be broken by an act of knowledge but always through an arbitrary act -- a resolution. \citep[quoting Mogernstern on p 129]{M_WWNMTA}
\end{quote}

Philosophically this means we cannot, reasonably, incorporate the probability of a Japanese earthquake into the modelling of asset prices, the epistemic problem, and it would be impossible to account for the irrationality of Nick Leeson in destroying Barings' Bank, the ontological problem.  Mathematically, these observations mean that we should not always assume that economies are ergodic, a point made by Paul Davidson \citep{D_RET}.  It does not mean that economics is not conformable to mathematics. When Feller discussed the application of measure theoretic probability he discussed  Pareto's law of income distribution, which he noted was dependent on the convenient assumption of ergodicity \citep[p 418]{F_TSP}.  Ergodicity simplifies the problem into well understood mathematics, but non-ergodic systems are also mathematically tractable, and are particularly important in the study of communications networks.  This point is made to stress that treating economic agents as cognitive beings does not undermine the status of economics as a mathematically based science, as Heilbroner argued, rather it highlights that economics requires its own mathematics developed in response to social, rather than physical, phenomena.

Beyond questioning the ability to foresee, or calculate,   Anscombe criticised consequentialism on the basis that acts obviously immoral, such as the execution of the innocent, could be justified either by the hope of good consequences or the fear of bad \citep[p 14]{A_MMP}.  Anscombe argued that to overcome this problem ethics should focus on the judgement of the \emph{agent} taking the action that has consequences, or Virtue Ethics.

Virtue Ethics, in the European tradition, is associated with Aristotle, in particular \emph{Nicomachean Ethics} in which virtues are the ``characteristics that enable individuals to live well in communities'' \citep[p 247]{P_CP}.  Aristotle's ethics do not distinguish reason and emotion, as Hume does, nor do they define absolute standards, rather virtue is a consequence of personal reflection \cite[pp 6--8]{vS_VEAP}.  This opens Virtue Ethics to the criticism that it cannot be codified into a set of rules that any person could apply to determine ethical action in any situation.  However, this criticism  assumes such a reduction is possible, and implicit in this is that the environment is stable, or ergodic.  The advantage of Virtue Ethics is that it can accommodate random change. 

Aristotle considered economics in Book V of \emph{Nicomachean Ethics}, in the context of a  broader discussion of the virtue of justice.  He distinguishes justice into two classes, distributive and directive (or corrective, restorative).  Distributive justice is concerned with the distribution of common goods by a central authority in proportion to the recipients' worth and is determined by equating  ratios, or Geometrical Proportions.  Directive justice  applies in cases where the parties are considered to be of equal worth, for example in commerce, in which case justice is determined by equating Arithmetic Proportion (\citep[p 41--42]{K_ENFC}, \citep[V.2.1130b30-1131a5]{Ethics}), and in this context justice is based on balanced reciprocity.

What is particularly significant in Aristotle's analysis is that he regarded the process of establishing both forms of justice as being one of establishing a mathematical equality. For example, in discussing  directive justice, Aristotle argues
\begin{quote}
 Now the judge restores equality; it is as though there were a line divided into unequal parts, and he took away that by
which the greater segment exceeds the half, and added it to the smaller segment. And when the whole has been
equally divided, then they say they have their own i.e. when they have got what is equal. The equal is intermediate between the greater and the lesser line according to arithmetical proportion \citep[V.2.1132a25-30]{Ethics}.
\end{quote}
This  is significant because at the start of \emph{Ethics} Aristotle distinguished mathematical and rhetorical arguments
\begin{quote}
Now fine and just actions, which political science investigates, admit of much variety and fluctuation of opinion, \ldots [therefore it is] foolish to accept probable reasoning from a mathematician and to demand from a rhetorician scientific proofs. \citep[I.3.1094b15--28]{Ethics}
\end{quote}

In \emph{Ethics}, Aristotle focused on \emph{oikonomike}, household, or moral, economy.  He also discussed political economy, \emph{chrematistike},  in \emph{Politics}, probably written after \emph{Ethics}.  The distinction of distributive and directive theories in  \emph{Ethics}, and the discussion of economic ideas in \emph{Politics}, has motivated a broad literature discussing what Aristotle's economic theories actually were and their relationship to modern economic theory \citep{P_ADRME}.  In the context of this paper we focus on directive justice, relevant to micro-economics, and do not consider distributive justice, which is principally concerned with macro-economics, and the subject of Rawls' \emph{A Theory of Justice}.

Out of Aristotle's discussion of market exchange, Scholastics developed the concept  of the `just price', which has been the subject of considerable modern debate.  For example, Raymond de Roover \citep{dR_CJP}, argues against viewing the just price in a Marxist, labour theory of value, sense but rather as the market price, in a neo-classical, liberal sense.  The basis of this paper's understanding of the just price comes from histories of science, such as Richard Hadden and Joel Kaye (\citep{H_OSM}, \citep{K_ENFC}), which have studied scholastic analysis of \emph{Ethics} in order to develop an understanding of the mathematisation of science.  The central point is that Scholastics, notably Albert the Great, realised that in approaching the ethics of fair exchange in a mathematical sense, Aristotle was ignoring the problem of incommensurability, that you can only compare like with like, and if this was done in the case of economics, it could be the case in physics.  Subsequent analysis of fair exchange by Thomas Aquinas and Pierre Jean Olivi is not just interesting from their economic perspective but also for the effect that they had on the development of western science. Hadden explicitly offers this as answering Needham's question as to why Chinese science did not develop as Western science did after the sixteenth century \citep[Chapter 4]{H_OSM}.

A definition of the just price from economic history that conforms to the one employed in this paper is provided by Fabio Monsalve \citep{M_SJPVCMP}, which is provided in the analysis of the late Scholastic, Juan de Lugo (1583--1660), rather than that of the generation Kaye and Hadden studied.
\begin{quote}
 Scholastic just price might be better understood if it is considered as a sort of intellectual construct: an ideal price that guarantees equality in exchange. It acts as a standard of measurement. Subsequently, it would
be possible to state whether the transaction was fair or unfair.

Just price is the price that guarantees equality in a transaction. It is the
amount of money that a thing is worth. If we only were to consider the just
price as an ideal price, there would be no major problems with the concept. The true problem arises when we try to determine if a real price is fair or unfair in a particular transaction. \citep[pp 6--7]{M_SJPVCMP}
\end{quote}

Monsalve goes on to point out, quoting Odd Langholm, that scholastic analysis was conducted in  a definite moral frame of reference, and so the `just price' ``could not refer indiscriminately to whatever price might be obtained in the market'' \citep[p 8, quoting Langholm]{M_SJPVCMP}.  This aspect was discussed in detail by the Scholastics prompted by a question `Whether the seller is bound to state the defects of the thing sold?'  posed by Thomas Aquinas \citep[ II, ii, qu. 77, art. 3, ad. 4]{Summa}.   Specifically Aquinas addresses the point
\begin{quote}
if the seller carry wheat to a place where wheat fetches a high price, knowing that many will come after him carrying wheat: because if the buyers knew this they would give a lower price.
\end{quote}
This question originates in Stoic philosophy and came to Aquinas through Cicero's \emph{De Officiis}.  In addressing the question, Aquinas offers the  observation  that
\begin{quote}
in the case cited, the goods are expected to be of less value at a future time, on account of the arrival of other merchants, which was not foreseen by the buyers. Wherefore the seller, since he sells his goods at the price actually offered him, does not seem to act contrary to justice through not stating what is going to happen. If however he were to do so, or if he lowered his price, it would be exceedingly virtuous on his part: although he does not seem to be bound to do this as a debt of justice.
\end{quote}
To Aquinas, the merchant  may {believe} that there are more grain shipments on the way, but does not {know}: the future is uncertain. On this basis, that the profit was not certain, the merchant can charge what the market will bear  based on the principle that
\begin{quote}
 because the just price of things is
   not fixed with mathematical precision, but depends on a kind of
   estimate, so that a slight addition or subtraction would not seem to
   destroy the equality of justice. 
\end{quote}
Kaye makes the point that Aquinas has separated the `just price', determined by divine law, from the `market price', established by men, Kaye explains the theological problem that Aquinas had stumbled upon, 
\begin{quote}
If just price equals common or market price and is divorced from individual judgement and direction, the individual's responsibility in economic activity is effectively eliminated. \citep[p 98]{K_ENFC}
\end{quote}
In addition,  Aquinas acknowledges that the just price is not a fixed, single point, but rather it defines a dynamic range of prices.  Providing the `market price' agreed by merchants is within this range, the merchant is not contravening natural law.  Aquinas has replaced {certainty} in knowledge with {estimation} \citep[pp 98--99]{K_ENFC}. 

Aquinas' stance was criticised by Pierre Jean Olivi, a leader of the `Spiritual Franciscans'.  The Spiritual Franciscans  argued that the vow of poverty meant monks should limit their use of property, not just not own property, \emph{usus pauper}, and Olivi was posthumously condemned as a heretic in 1326, hindering the transmission of his philosophy.  The Franciscans, unlike the empirical rationalist Dominicans, were fideists and this philosophical approach  meant that Olivi argued that the metaphysical \emph{probability} of more grain arriving, giving the merchant excessive profits,  had a certain reality, which Aquinas was ignoring by focusing on the `physical reality' of the prices being offered in the market \citep[p 121]{K_ENFC}.  Olivi said
\begin{quote}
 The judgement of the value of a thing in exchange seldom or never can be made except through conjecture or probable opinion, and not so precisely, or as if understood and measured by one invisible point, but rather as a fitting latitude within which the diverse judgements of men will differ in estimation \citep[p 124]{K_ENFC}.
\end{quote}
This distinction is essential in demarcating the just price from the market price \citep[Section 3.2.1]{M_SJPVCMP}.  Olivi  applies this approach to the question of loans, as Kaye explains
\begin{quote}
 if someone intends to invest his money in trade or profit, and instead, out of charity, lends the money to a friend in need, can he expect back from his friend not only the sum lent but in addition the profit he lost in not investing in trade?  Olivi's answer to this question was an unqualified yes: the borrower was responsible for indemnifying the lender for his loss of ``probable profit'', and for restoring a ``probable equivalence''. (\citep[p 119]{K_ENFC}, also \citep[pp 265--267]{F_SC})
\end{quote}
Olivi introduces the idea that market exchange is about equating expectations, not concrete values.

In Aristotle \citep[V.2.1132a25-30]{Ethics} we can see the argument that an asset price is a point on a line, just as in the simple description of a martingale price given above. Out of the Scholastic analysis of the `just price' undertaken in the context of Virtue Ethics, two key features of the FTAP, that a price is an expectation and that in actual, as opposed to ideal, markets a price cannot be precisely defined, seem to have been established. However,these observations were philosophical, not mathematical, it is in the seventeenth century that  the synthesis of ethics and commercial arithmetic  produces modern mathematics that suggest the mathematical FTAP is equivalent to these ethical observations.

\section{Finance and the development of mathematics}

Medieval European merchants had to contend with at least two factors of the economy: prohibitions on usury and the heterogeneity of currency.   

Usury derives from the Latin \emph{usus} meaning `use', and referred to the charging of a fee for the use of money.  Interest comes from the Latin \emph{interesse}, and originated in the Roman legal codes as the compensation someone was paid  if they suffered a loss as a result of a contract being broken \citep[p 73]{HIR}.  Shortly after 1200 the theologian, Peter the Chanter, argued that ``a buyer or seller may be excused from usury if he exposes himself to the risk of receiving more or less \citep[pp 263--264]{F_SC} and some 40 years later,  Alanus Anglicus,  determined that \emph{turpe lucrum}, the `shameful gain' of ''asking for more than what was given'', did not exist if the future price of the good was uncertain in the mind of the merchant {\citep[p 41]{ETBAS}}.  These judgements were the basis of Aquinas' and Olivi's later analysis of the just price.

The basic financial instrument at this time was the census  that originated when Carolingian monasteries guaranteed donors a fixed regular income in exchange for a bequest of land.   Censii evolved to be written on the back of assets as diverse as a grand estate or a craftsman's labour and the  contracts developed such that a borrower would receive a lump sum secured against the future cash-flow from an asset, \emph{rente \`{a} prix d'argent}. Since the income stream from the assets underlying a census was uncertain while the  payments from the census were  fixed, there was an implicit swap involved in the contract.  

Censii had many features of contemporary asset backed securities, the triple, or German, contract (\emph{contractus trinus}), developed to fund long distance trade, had the features of modern structured finance.  It involved a loan to fund the venture (the first contract); the transformation of the variable return of the venture into fixed cash-flow (the second contract); and an insurance contract to guarantee the fixed payment (the third contract), a  Credit Default Swap.  This contract was declared illicit in 1586 on the basis that the lender received a riskless return [\emph{Detestabilia avarita}].

 William Goetzman explains that as a consequence of the multitude of currencies, European  medieval merchants, 
\begin{quote}
operated in a world of complete relativism. With no central government, no dominant currency, and even competing faiths and heresies, value is expressed quite abstractly only in a set of relative relations to other items.  {\citep{G_FFR}}
\end{quote}
Alfred Crosby  comments that this heterogeneity of currency and the complexity of the contracts employed meant that there was an  ``abstraction of Western merchants' scale of value''  {\citep[p 72]{C_MR}} and that ``Westerners were obsessed with what they could not hold on to, money'' and  as a consequence ``no people were more obsessed with counting and counting and counting''  {\citep[p 74]{C_MR}}. The 1202 publication of Fibonacci's \emph{Liber Abaci} enabled merchants to navigate their way through the complex contracts in a sea of shifting prices with an eye on the usury laws. 

The most obvious innovation by Fibonacci was the introduction of  Indian numbers  and   meant that merchants could write down their calculation method, the algorithm, which could then be copied, modified and improved by others: knowledge could be distributed and improved more easily.  \emph{Abacco} or \emph{rekoning} schools sprang up throughout Europe teaching apprentice merchants the material  in the book.    The impact of these abacco schools  was enormous, algebra became an important tool used by the large and influential community of Europeans and would provide the reservoir of mathematicians on which the scientific developments of the seventeenth century were built  (\citep[Chapter 1]{H_OSM},  \citep[Introduction]{LA}, \citep{H_AT} ).

 The medieval  approach to probability followed  Aristotle, who simply classified events into three types: certain events determined by specific causes; probable events that usually happened; and unpredictable events and games of chance were unpredictable and not amenable to science \citep[p 30]{H_HPS}, a classification that is not that dissimilar to Knight's or Keynes' in the twentieth-century. While medieval merchants became adept at pricing aleatory contracts and theologians considered problems of epistemology in casuistry, the concept of mathematical probability, of measuring chance, did not emerge until the mid-sixteenth century with Cardano's \emph{Liber de Ludo Alea}, wrtten, probably after some decades of thought, in 1564. Ian Hacking has remarked that the emergence of  the concept of absolute chance was late but the identification of mathematical probability precedes both  Descartes' introduction of absolute space (Cartesian co-ordinates) and Newton's of absolute time.

Up until the 1950s, and a re-assessment of  his work by \O{}ystein Ore \citep{O_CGS}, Cardano's contribution to probability theory had been widely ignored.  In the context  frequentist interpretations of probability, that dominated the nineteenth and early twentieth centuries, it was seen as incoherent.  More recently, David Bellhouse has  re-evaluated the \emph{Liber} looking at it as a humanist philosophical text, not as a mathematical document,  based on the fact that Cardano, himself, did not list it as one of his mathematical works.  Bellhouse's hypothesis is that in the \emph{Liber} Cardano is trying to establish under what grounds  gambling can be considered ethical, and as a Renaissance  scholar he would have done this in the humanist manner, in the context of \emph{Nicomachean Ethics} \citep{B_DLLA}.

  Cardano, just like the thirteenth century Scholastics,  latches on to the idea that justice  is equivalent to equality\label{CarEq}, and observes that 
\begin{quote}
 The most fundamental principle of all in gambling is simply equal conditions, e.g., of opponents, of bystanders, of money, of situation, of the dice box, and of the die itself. To the extent to which you depart from that equality, if it is in your opponent's favour, you are a fool, and if in your own, you are unjust. {\citep[quoting from Chapter 6 of the \textit{Liber}]{B_DLLA} } 
\end{quote}
Based on this thinking, Cardano realised that in dice games `equality' was established by counting the ways a player could win and comparing that number to the ways a player would lose.  On this basis the `chance' of winning could be deduced, and if  the stakes did not match the chances, the gamble was unjust.  Summarising his findings  he states,  ``a just gamble is one between willing and knowledgeable players'' and almost immediately after coming to these fundamental conclusions, Cardano observes that 
\begin{quote}
 These facts contribute a great deal to understanding but hardly anything to practical play {\citep[p 58 quoting from Chapter 9 of the \textit{Liber}]{GGG}}
\end{quote}
 since the gambler, even through mathematics, could not predict the future outcome of the die cast.  Both these aphorisms are as applicable today as they were to Cardano.

Understanding the mathematics of probability did not tell Cardano what would happen in the future, rather it helped him understand how he should act in an uncertain world. In particular,  Cardano was using mathematics to decide what constituted a just act, not as an instrument to aid his gambling.  He is approaching the problem as Aristotle might: the virtuous person uses their reason to identify the best course of action.  

One problem Cardano considered was the so-called \emph{Problem of Points}.  This problem appears in Pacioli's abacco text \emph{Summa de arithmetica, geometria, proportioni et proportionalit\`{a}} (`Work on arithmetic, geometry and proportion') and is based on the following situation:
\begin{quote}
 Two players, $F$ and $P$, are playing a  game based on a sequence of rounds, and each round consists of, for example, the tossing of a fair coin.  The winner of the game is the player who is the first to win 7 rounds, and they will win 80 francs.
\end{quote}
  The \emph{Problem of Points} is how the 80 francs should be split if the game is forced to end after $P$ has won 5 rounds while $F$ has won 4. 
Edith Dudley  Sylla  notes that the \emph{Problem} comes from the abacco tradition of using `stories' to give examples of how to solve problems in commercial arithmetic, a `Babylonian' rather than `Greek' approach to mathematics (e.g. \citep[pp 44--45]{F_CPL} and in economics, \citep{D_ABT} ). In this case the  \emph{Problem of Points}, the story represents the case of how the capital tied up in a business partnership should be divided up if the venture has to finish prematurely \citep{S_BECM}.  

Pacioli's solution was statistical, the pot should be split 5:4.  Cardano realised this was absurd since it would give a manifestly unfair result if the game ended after one round out of a hundred or when $F$ had 99 wins to $P$'s 90.  Cardano makes the point that the correct solution would be arrived at by considering what would happen in the future, it had to be forward-looking, in particular it had to account for  what `paths' the game would follow.  Despite this insight, Cardano's solution was still wrong, and the correct solution was provided by Pascal and Fermat in their correspondence of 1654.

The Pascal-Fermat solution to the \emph{Problem of Points} is widely regarded as the starting point of mathematical probability.  The pair (it is not known exactly who) realised that when Cardano calculated that $P$ could win the pot if the game followed the path $PP$ (i.e. $P$ wins and $P$ wins again)  actually represented four paths, $PPPP$, $PPPF$, $PPFP$, $PPFF$; it was the players `choice' that the game ended after $PP$, not a feature of the game itself; an early example of disentangling  behaviour from problem structure.  Calculating the proportion of winning paths would come down to using the Arithmetic, or Pascal's, Triangle -- the Binomial distribution --  and the value of the game was simply the expectation of the game given where on the tree of paths the players where, as a martingale.  Essentially,  Pascal and Fermat established what would today be recognised as the Cox-Ross-Rubenstein formula, based on the Binomial distribution, for pricing a digital call option \citep{CRR_OP}.

%The claim that probability theory originates with the solution to pricing an option in the modern manner is open to the criticism that  the key feature of the Cox-Ross-Rubenstein approach is the use of risk-neutral probabilities, and it is `whiggish' to identify the Pascal-Fermat approach with Cox-Ross-Rubenstein.
The Pascal--Fermat correspondence was private, the first textbook on probability was written by Christiaan Huygens in 1656.  Huygens had visited Paris in late 1655 and had been told of the \emph{Problem of Points}, but not of its solution (\citep[p 111]{GGG}, \citep[p 67]{H_HPS}), and on his return to the Netherlands he solved the problem for himself and produced the first treatise on mathematical probability, \emph{Van Rekeningh in Speelen van Geluck} (`On the Reckoning at Games of Chance') which would appear in van Schooten's \emph{Exercitatonium Mathematicarum},  a  textbook, as \emph{De Ratiociniis in Ludo Aleae} in 1657.

In \emph{Van Rekeningh} Huygens starts with, what is essentially, an axiom, 
\begin{quote}
I take as fundamental for such [fair] games that the chance to gain something is worth so much that, if one had it, one could get the same in a fair game, that is a game in which nobody stands to lose.\ent{H_HPS}{p 69}
\end{quote}
Probability is defined by equating future gain with present value, as a martingale measure, in the context of `fair' games, in the absence of arbitrage.

The case that Huygens was working in the context of Virtue Ethics is enhanced by recognising  the difficulty he had in translating \emph{Van Rekeningh} into Latin \citep[pp 93--94]{EoP}.  Huygens struggled to translate the Dutch word \emph{kans} (`chance', `lot'), which would normally be translated as \emph{sors}, and eventually he, or his editor van Schooten, chose \emph{expectatio}, giving the  English term `expectation' (in the mathematical sense).  However, Huygens had considered using the Latin word \emph{spes} \citep[p 95]{EoP} which was the  term for the virtue `Hope'. In French, \emph{esp\'{e}rance} is used when referring to mathematical expectation, reflecting this debate. The  Dutch, who following Stevin's focus on teaching mathematics in the vernacular, use their own terms in mathematics, in this case the  equivalent is \emph{verwachting}: hope, promise, expectation, forecast, prognosis.

Fifteen years after writing \emph{Van Rekeningh}, Christiaan Huygens made further developments in probability, particularly in conditional probability associated with dependent events, in addressing the problem of pricing life-annuities in a correspondence with his brother, Lodewijk, who was a government official in the Dutch Republic.  The problem was solved by the leader of the Republic at the time, Jan De Witt, employing the law of one price.  At the same time, books describing gambling emerged in the context of Louis XIV's \emph{appartements du roi}, thrice weekly gambling events that have been described as a `symbolic activity' not unlike \emph{potlach} ceremonies \citep[pp 31-42]{K_ESC}.  This important social activity provided further stimulus in the development of objective, or frequentist,  probability.

Around 1684 James Bernoulli, who would emerge as one of the most important mathematicians of his generation alongside Newton and Leibniz, began working on problems in probability and between 1700 and his death in 1705, he worked on  \emph{Ars Conjectandi} (`The Art of Conjecturing'), which was published posthumously in 1713.   The \emph{Ars} is made up of four parts, a commentary on Huygens' \emph{Van Rekeningh}, original work on calculating permutations and combinations, applications of these ideas to games of chance and finally the application of the ideas to ``civil, moral and economic affairs'' \citep[p 224]{H_HPS}. 

While the first three sections of the \emph{Ars}  are un-controversial, the final section is both the most significant and has proved problematic.  Bernoulli, having discussed objective probability at length introduces the epistemic, or subjective, definition of probability as ``a degree of certainty''.  Anders Hald notes that this is ``revolutionary'' in that Bernoulli is applying mathematics to propositions, not just events, and that he believed subjective probability was universally applicable, though his analysis would only accommodate situations of stable relative frequencies \citep[p 225]{H_HPS}.  This section of the \emph{Ars} is significant in that it introduces what would become known as the `law of large numbers', which can be summarised as collecting a large amount of data will improve the accuracy of an observation -- providing the system was stationary \citep[p 225]{H_HPS}.  The section is problematic  because Bernoulli considered situations where the sum of probabilities could be greater than one\ent{S_CAT}{p 27}.  This is impossible if probability was calculated as relative frequency, but a feature of calculating the value of commercial contracts, or when arbitrage is present, as in a Dutch book.

Sylla, who has compared Bernoulli's work to that of  Huygens' and other contemporaries, de Witt and de Moivre, in the process of translating the \emph{Ars}, has  concluded that
\begin{quote}
 equity among associates or partners rather than probabilities in the sense of relative frequencies provided the foundation for the earliest mathematical probability theory.\ent{S_CAT}{p 13}
\end{quote}
%Sylla realised that Bernoulli's statement that 
%\begin{quote}
%the fundamental principle of the whole art \ldots: Anyone may expect, or should be said to expect, just as much as he will obtain without fail\ent{S_CAT}{p 28}
%\end{quote}
%was the same as Huygens' introductory axiom, and was similar to  examples from de Witt and de Moivre.  She  concluded that
%\begin{quote}
% equity among associates or partners rather than probabilities in the sense of relative frequencies provided the foundation for the earliest mathematical probability theory.\ent{S_CAT}{p 13}
%\end{quote}
and that
\begin{quote}
 While traditional histories of mathematical probability start with Pierre Fermat, Pascal and Huygens because they give what are from the modern point of view correct frequentist solutions to the problems of division and expectations in games of chance \ldots the foundations of Huygens' method (\ldots) was not chance (frequentist probability ), but rather \emph{sors} (expectation) in so far as it was involved in implicit contracts and the just treatment of partners.\ent{S_CAT}{p 28}
\end{quote}
Seventeenth century (financial) mathematicians would not have used the `natural' measure based on observed relative frequencies, but would have  imagined a world of `ethical' martingale measures that established equality between present price and future value.

Sylla also makes some comments about what we would now regard as incomplete markets when she observes that  \emph{The Port Royal Logic},  a significant influence on Pascal, notes that ``because the house takes part of the stakes, lotteries are manifestly unfair'' and seventeenth century mathematicians
\begin{quote}
 assumed that gamesters were gentlemen or --women who put up a stake to play a game and divided the resulting  pot of money at the end of the game according to the previously agreed upon rules.  There was no `house' or government that skimmed part of the players' wagers off the top.\ent{S_BECM}{p 327}
\end{quote}
The implication is that seventeenth century mathematicians, Pascal at least, would have seen transaction costs as being unethical and, while utopian, frictionless markets  are `fair'  and so the proper subject of study.

When comparing the contemporary FTAP  with late-seventeenth century approaches to valuing contracts, associated with the emergence of mathematical probability up until Bernoulli's \emph{Ars Conjectandi}, there appear to be striking similarities.  The Pascal-Fermat solution to the Problem of Points can be identified with the Cox-Ross-Rubenstein simplified approach to option pricing, formulated in the mid-1970s.  Cardano, Huygens and Bernoulli all associated fair games with mathematical expectations, just as the FTAP associates the corresponding concepts of no-arbitrage with martingale measures.  Finally there was a recognition that transaction costs hinder the mathematical analysis of fair prices, just as the FTAP states that unique prices do not exist in incomplete markets.

Sean Mahony warns of the perils of Whig history  \citep[xiv--xv]{M_MCPF}: histories of mathematics re-write what seventeenth century mathematicians wrote using modern notation to demonstrate that they thought as we think, the immutability of mathematics.   We do not claim that Huygens approached the problem of pricing contracts in the same way that Black, Scholes and Merton, or Heath, Jarrow and Morton have. There is no evidence to suggest that when  Black and Scholes stated that ``it should not be possible to make sure profits'', the `no arbitrage' condition, they were  conforming to Anglicus' argument that  \emph{turpe lucrum} did not exist if the future  was uncertain.  

The solution to the conundrum of the apparent equivalence of the FTAP and seventeenth century arguments could lie in the nature of mathematical inference.  Inferential  statistics can be described as seeking to obtain the best estimate for a parameter given a set of experimental outcomes, and this involves giving different weights to the outcomes of different experiments.  In classical statistics this is achieved through observing frequencies, in subjective probability it is made through judgement.  In the context  of finance, where the objective is to estimate the current price based on possible future prices,  the experimental outcomes of which only one is realised, the inference should be made on  the basis of fairness.  This observation does not solve the problem of epistemic or ontological uncertainty, which is a feature of frequentist and subjective frameworks, rather it changes the basis of economic analysis.

In the final two Sections we shall investigate why the insight  that fairness is important in calculating the value of financial contracts was lost in the nineteenth century and what implications it might have on contemporary financial mathematics.

\section{Uncertainty, scarcity and reciprocity}

The first academic paper by the mathematician Kolmogorov had been in economic history, when he submitted it he was told that ``You have supplied one proof of your thesis, and in the mathematics that you study this would perhaps suffice, but we historians prefer to have at least ten proofs.''\ent{S_KLCA}{}, this captures the difficulty of giving  a rigorous answer to the question posed at the end of the previous section.  However, inspired by Moses ben Maimon (Maimonides) we attempt to provide an explanation. Ben Maimon argued in  his  \emph{Guide for the Perplexed}, written in 1190 and an influence on Aquinas, that God's punishment after the Fall of Man  was not so much about scarcity but uncertainty. In the Garden of Eden humans had perfect knowledge, which was lost with the Fall, and it is the loss of this  knowledge which is at the root of suffering: if we know what will happen we can manage scarcity {\citep{P_LOMP}}. In response to this observation, we present a narrative of history that suggests economic thought switches from being concerned with scarcity (characterised by famine and pandemics) to  uncertainty (characterised by war), and this provides an answer to the question.  

The earliest Greek text that addresses economic questions, \emph{Works and Days}, was written by the  poet Hesiod \citep{G_AH} around 750 BCE, pre-dating the scientist-philosopher Thales buy some one hundred years.  The poem starts with the myth of Pandora, relating the strife of mankind to the introduction of technology and goes on to identify the  role that scarcity has in determining human behaviour; the gods make life a struggle for mankind, and so we have to work hard.  The archaeological record for the Aegean at this time  suggests that climate change led to  a long running famine  and there was migration from the main city-states into new areas for farming and the poem describes how Hesiod's father had been a farmer and merchant in north-western Anatolia, but the recession had forced him  to move to Boeotia in modern Greece.

Plato discussed economic concepts in \emph{The Republic}, where he argues (through Socrates) that states arise ``out of the needs of mankind; no one is self-sufficing, but all of us have many wants.'' \citep[II, 369]{Republic}.  According to Schumpeter, 
\begin{quote}
 Plato's [state was] conceived for a small and, so far as possible, \emph{constant} number of citizens.  As \emph{stationary} as its population was to be its wealth.  All economic and non-economic activity was strictly regulated \citep[p 52, my italics]{S_HEA}
\end{quote}
 \emph{The Republic}, which has been described as communist/fascist \citep[p 52]{S_HEA}, was written in the shadow of Athens' defeat by Sparta, \emph{Nicomachean Ethics}, with its emphasis on reciprocity, was probably written  around 330 BCE  when Aristotle was connected to  the court of the ascendant Macedonia.

The Roman legal codes placed an emphasis on property rights and stated that the `just price' of a good was one that had been agreed, voluntarily, between any buyer and seller  \citep[p 31]{ETBAS}.  The decline of Rome in the west was accompanied by population falls and labour shortages from about the second century \citep[p 42]{P_EHME} and following the sack of Rome by the Goths in 410 CE, Augustine of Hippo wrote  \emph{De Civitate Dei contra Paganos} (`The City of God against the Pagans').  The book
\begin{quote}
 Censures the pagans, who attributed the calamities of the world, and especially the recent sack of Rome by the Goths, to the Christian religion, and its prohibition of the worship of the gods\ent{CoG}{Sub-title}
\end{quote}
The pagans were arguing that the destruction of Rome, a disastrous event that impoverished the city, was a consequence of Christianity. Augustine countered, and laid the foundations of the Catholic Church,  by offering a Platonic argument that the corporeal world was an unnatural state for the Christians, who would eventually return to The City of God.  To counter Pelagian and Donatist heresies Augustine  argued against free-will and while he agreed with the Stoics about the  uselessness of divination and astrology in predicting the future,  he disagreed with the pagan view that God had no foreknowledge 
\begin{quote}
 God knows all things before they come to pass, and that we do by our free will whatsoever we know and feel to be done by us only because we will it . \ldots  all wills also are subject [to God], since they have no power except what He has bestowed upon them.\ent{CoG}{The City of God, Book V, Chapter 9 }
\end{quote}
In his book \emph{On free choice of the will} he relies on mathematical analogies to convince his audience that there are transcendental truths
\begin{quote}
For those inquirers to whom God has given the ability \ldots [can see that] the order and truth of numbers has nothing to do with the senses of the body, but that it does exist, complete and immutable, and can be seen in common by everyone who uses reason\ent{A_FCW}{p 46}
\end{quote}
Augustine is incorporating the immutability and indubitablity of mathematics as Catholic dogma.

Between 1000--1300 CE Europe underwent a period of economic expansion that  saw the population double while the amount of coin in circulation is thought to have increased six-fold. Driving this  expansion was a rise in temperature that led to greater agricultural productivity, population growth and after the late twelfth century,  an expansion of commerce, centred on the Mediterranean,  accompanied by an urban explosion.  This great expansion of Latin Christianity  went into reverse by the start of the fourteenth century \citep[p 122--123]{P_EHME}.    Between 1315 and 1322 abnormally cold winters were separated by abnormally wet or dry summers, harvests failed  leading to the Great Famine in which  about 10\% of the population died in some urban areas.  These famines were followed in 1347 by the Black Death, and by 1350  the plague had spread throughout Europe, causing the death of, in places, a third of the population, hitting the poor and urban communities particularly hard.  The plague was not a single event, but reappeared, and the successive calamities of the 1320s, 1350s and then 1360s would have traumatised society. The influence of the Church suffered  as clerics were disproportionately infected by plague, since they cared for the sick, and it proved unable to protect people from the disease.  The Scholastics, using either faith or reason, were unable to explain what was happening around them, and in the words of  William Bouwsma, they failed to ``give to life a measure of reliability and thus reduce, even if it cannot altogether abolish, life's ultimate and terrifying uncertainties'' {\citep[p 58 quoting \emph{Anxiety and the Formation of Early Modern Culture}]{C_MR}. The agricultural collapse in northern Europe and disruptive wars led to a collapse in commerce, particularly hard hit were the Flemish towns that had emerged as centres of trade following Philip IV the Fair's suppression of the Champagne Fairs between 1305 and 1309.

After Latin Christianity lost its empire in the Eastern Mediterranean the Crusades were replaced by dynastic conflicts.    While England and France had strong monarchies, fighting each other in the Hundred Years War between 1337 and 1453, the Holy Roman Empire and Italy became fragmented.  Merchant cities on the Atlantic and Baltic coasts  exerted their independence from German princes, who themselves had vast estates and were capable of challenging the Emperor. The cities of northern Italy became dominated by \emph{signori} while the Papacy and the Kingdom of Naples became  pawns in the conflict between the French and Spanish/Hapsburg monarchies.

  The primary concern of the autocrats was the maintenance of their power  based on armies manned by mercenaries that needed to be paid: \emph{Pecunia nervus belli}, `money is the sinew of war'. The subtleties of Olivi's economic analysis would be replaced by demand for scarce gold and silver.

By the start of the sixteenth century a rise in population  \citep[p 144]{P_EHME} suggests that the the scarcity associated with the previous centuries had ended and there was a return to uncertainty.   John Julius Norwich describes the situation of Cardano's youth
\begin{quote}
In and around Milan the fighting had hardly ever stopped; there must have been many Milanese who, on waking in the morning, found it difficult to remember whether they owed allegiance to the Sforzas, the Emperor or the King of France.\ent{N_MS}{p 273}
\end{quote}

The anxiety of the fourteenth and fifteenth centuries led to a   revival of strict Augustinian theology advocated by Jean Calvin.  The reformed church  proved popular amongst mercantile communities that were growing as  wealth and power moved from the Mediterranean to the Atlantic following the discovery of the Americas. Confessional wars erupted, in France, the Netherlands, Central Europe and the British Isles. The  Dutch Wars, initially stimulated by high Hapsburg taxation resulted in the rebels paying a higher tax burden \citep[p 11]{FHN} but  significant developments in public finance  enabled the Dutch to successfully fund their rebellion \citep[Chapter 2]{FHN}.  Key in this process was Simon Stevin, the abacco trained mathematician who established the Dutch Mathematical School that inspired Descartes. The Thirty Years War, characterised by \emph{bellum se ipse alet}, `let war pay for itself',  resulted in widespread devastation, but the `Gothic atrocity narrative' that emerged during Romanticism has been re-evaluated in the twentieth century, and  contemporary historians argue that ``Sudden changes in fortune became a defining characteristic of the conflict'' \cite[p 845]{W_ET}.  The peace treaties that ended the conflicts established the principle of Westphalian sovereignty, that endorsed homogeneity within a state.

In London: 1665 saw Plague; 1666 the Great Fire; 1672 The Great Stop of the Exchequer;   1673 an Act of Common Council that looked to put an end to ``usurious contracts, false Chevelance, and other crafty deceits'' {\citep[p 83]{M_OEFM}}; an unsuccessful revolution in 1685, a successful one in 1688; there was a stock-market boom in the early 1690s, with around 40\% of the trades between 1692 and 1695 being in stock options \citep[p 24--30]{M_OEFM}; 1694 saw the \emph{The Million Adventure Lottery} and the creation of the \emph{Bank of England}; in 1696 was the re-coining, precipitating what has been described as ``the gravest economic crisis of the century''  {\citep[p 56]{M_OEFM}} that was partially resolved by Isaac Newton; 1697 saw the first legislation ``To Restrain the number and ill Practice of Brokers and stock-jobbers'' being passed.  The  turbulence of these thirty-two years see the creation of the City of London that is familiar today and laid the foundations of Britain's Empires over the next three hundred years.  The 1707 Acts of Union between the English and Scottish parliaments created a unified state in the British Isles, while in France, the 1685 the Revocation of the Edict of Nantes ended a century of a Protestant `state' within a predominantly Catholic France.

The frequentist approach  began to dominate the mathematical treatment of probability following the claimed `defeat', or `taming',  of chance by mathematics  with the publication of Montmort's \emph{Essay d'Analyse sur les Jeux de Hazard} (`Analytical Essay on  Games of Chance') of 1708  and De Moivre's \emph{De Mensura Sortis} (`The Measurement of Chance'), of 1711 developed in \emph{The Doctrine of Chances} of 1718 \citep{B_BF}. All these texts were developed more in the context of gaming than in the analysis of commercial contracts, which had been the focus of the work of Pascal, Huygens and Bernoulli. \emph{The Doctrine} was the more influential, introducing the Central Limit Theorem, that  independent random variables will be Normally distributed, a result that can be used to argue that asset prices should be log-Normally distributed, as in the Black-Scholes-Merton model.  By 1735 it was being argued that the achievements of the seventeenth century probabilists from Huygens to  de Moivre superseded the classical approach to probability, dividing events into the certain, probable and unpredictable,  in that it measured probability and changed the status of the `unpredictable' \citep[p 578]{B_BF}.  

It is remarkable that the development of mathematical probability was undertaken almost exclusively by Augustinians: Pascal was a Jansenist; Huygens, Bernoulli and de Moivre were Calvinists;  Montmort had been trained to be an Augustinian but renounced orders to marry.  This observation is compounded by the facts that Newton was an Arian/Anglican and Leibniz a Lutheran and neither did significant work in probability, Fermat was a Catholic living in the mixed Calvinist/Catholic city of Toulouse. As Augustinians the probabilists all believed in God's pre-destination  and omniscience: they denied the existence of randomness, events  were unpredictable because man could not understand God's intentions.  Over the course of the next hundred years the implicit determinism of the Augustinians became a standard feature of Western science, being codified by Laplace in the 1820s (\citep{D_CPE}, \citep[pp 387-390]{R_UT}).

From the mid-eighteenth century Romanticism, emphasising an individual genius' emotional reaction to nature, emerged to eclipse the Classical rationalism of the Enlightenment.     By the end of the  century, the rural cleric and mathematician,  Thomas Malthus captured the   anxiety  of the  English gentry following the Terror of the French Revolution  in  \emph{An Essay on the Principle of Population} that focused on scarcity.  
At the height of Romanticism in 1836 John Stuart Mill defined political economy as being 
\begin{quote}
  concerned with [man] solely as a being who desires to possess wealth, and who is capable of judging of the comparative efficacy of means for obtaining that end. \citep{M_DPE}
\end{quote}
and  defended Malthus in \emph{Principles of Political Economy} of 1848, written at a time when Europe was struck by the Cholera pandemic of 1829--1851 and the famines of 1845--1851 and  Alfred, Lord Tennyson, was describing nature as ``red in tooth and claw''. Alfred Marshall, synthesised Mill's approach to economics  with Darwinian  metaphors of competition (\citep[10.1]{B_HMEA}, \citep{T_AMEB}) to lay the foundations of  neo-classical economics.  Marshall's 1890 definition of economics  would be paired down by Lionel Robbins in 1932, towards the end of the Great Depression, as ``the science which studies human behaviour as a relationship between ends and scarce means which have alternative uses''.  

A consequence of Romanticism was the emergence of Nationalism.   The 1820s saw a re-birth of Scottish identity within the Union and the Greek revolt against Turkey while the 1840s saw the Italian Wars of Independence from the Austrian Empire and the Young Ireland movement.  The super-states that emerged after the Peace of Westphalia began to fragment, not through external interference but internal strife.  Outside Europe, the United States had chosen Federalism in 1787, when the nation faced an uncertain future, but through the next century, states' rights were increasingly asserted, particularly in economic affairs.

In the years before the First World War economic policy, particularly interest rates that governed credit,  was dominated by the gold-standard, while after the Second World War it was Bretton-Woods.  The collapse of Bretton-Woods in August 1971  had an immediate impact on interest-rates.  In the 27 years between 1945 and autumn 1972, the Bank of England changed its lending rate 43 times, in the 27 years after 1972, it changed them 223 times.  A key economic factor had gone from being fairly stable to being a random process.  Similarly, as the US dollar fell in value, price setting mechanisms in commodities, notably oil, collapsed.  It was in this environment that financial derivatives made a re-appearance after having been dormant for half a century.  The collapse of Bretton-Woods was an inevitable  consequence of the non-ergodic nature of global society; the ``30 glorious years of capitalism'' between 1948--1977 was, to a large degree, a consequence of Western hegemony, it was replaced by  ``30 glorious years of capitalism'' in the East.

I suggest that when faced with scarcity, society responds by fragmenting into elements that compete for scarce resources. Alternatively,  when society is challenged by uncertainty it turns to communality, seeking to diversify risks.  On this basis, the rise of expected utility maximisation as dealing with scarcity in a Romantic context of the individual genius struggling against nature and within a framework of  stable chances, can be explained. I contend that since the `Nixon shock' society has been  more focused on uncertainty than scarcity and is struggling to shift the economic paradigm in response to this change in the economic  environment.  

This is captured in Arjun Appadurai argument that the leading agents in modern finance 
\begin{quote}
 believe in their capacity to channel the workings of chance to win in the games dominated by cultures of control \ldots [they] are not those who wish to ``tame chance'' but those who wish to use chance to animate the otherwise deterministic play of risk [quantifiable uncertainty]''. \citep[p 533-534]{A_GFM}
\end{quote}
These observations conform to the definition of a `speculator' offered by  Reuven and Gabrielle Brenner: a speculator makes a bet  on a mis-pricing.  They point out that this definition  explains why speculators are regarded as socially questionable: they have opinions that are explicitly at odds with the consensus (\citep[p 91]{B_GS}, see also \citep[p 394] {BS_DR}).  In comparison, gamblers will bet on an outcome: an ace will be drawn; such-and-such a horse will win against this field in these conditions; or Company Z will outperform Company Y over the next year.  Investors do not speculate or gamble, they defer income, `saving for a rainy day', wishing to minimise uncertainty at the cost of potential profits.

Speculation in  modern finance, as Daniel Beunza and David Stark observe,  is about understanding the relationship between different assets and   ``to be opportunistic you must  you must be principled, i.e. you must commit to an evaluative metric'' \citep[ p 372]{BS_TTSTA}.  This explains why modern finance, just as Goetzman's medieval merchant,  relies on mathematics,  ``Mathematicians do not study objects, but the relations between objects''  {\citep[p 22]{P_SH}}.

The speculator has been a feature of the modern markets ever since they were established in the seventeenth century.  In 1719 Daniel Defoe   described stock-jobbing  in  \emph{The Anatomy of Exchange Alley}  as
\begin{quote}
a trade founded in fraud, born of deceit, and nourished by trick, cheat, wheedle, forgeries, falsehoods, and all sorts of delusions; coining false news, this way good, this way bad; whispering imaginary terrors, frights hopes, expectations, and then preying upon the weakness of those whose imaginations they have wrought upon {\citep[ p 290 quoting Defoe]{EHFE}}
\end{quote}
The process described is speculation because the stock-jobbers are not betting on outcomes, rather they are attempting to change the expectations of others, create a mis-pricing that they can then exploit.

The philosophical antecedents of the modern hedge fund manager, betting against determinism, could be even further back, in the medieval Franciscans  such as Pierre Jean Olivi  and  John Duns Scotus.  While the Dominican, empirical rationalist, Aquinas argued that knowledge rested on reason and revelation, Scotus argued that reason could not always be relied upon: there was no true knowledge of anything  apart from theology founded on faith.  While Aquinas argued that God could be understood by rational examination of nature, Scotus believed this placed unjustifiable restrictions on God, who could interfere with nature at will: God, and nature, could be capricious \citep[p 127]{L_MT}.

Appadurai was motivated to study finance  by Marcel Mauss' essay \emph{Le Don} (`The Gift'), exploring the moral force behind reciprocity in primitive and archaic societies.  Appadurai notes that the speculator, as well as the gambler and investor, is  ``betting on the obligation of return'' \citep[p 535]{A_GFM}.  The role of this balanced reciprocity in finance can be seen as an axiom in that  it lays the foundation for subsequent analysis, it can also be seen as a simplifying assumption: if the future is uncertain what mechanism ensures that agreements will be honoured. David Graeber also recognises the fundamental position reciprocity has in finance \citep{G_D}, but where as Appadurai recognises the importance of reciprocity in the presence of uncertainty,  Graeber essentially ignores the problem of in-determinism in his analysis that ends with the conclusion that ``we don't `all' have to pay our debts'' \citep[p 391]{G_D}.

Aristotle's \emph{Nicomachean Ethics} is concerned with how an individual can live as part of a community and he  saw reciprocity in exchange as being important in binding society together (\citep[p 51]{K_ENFC}, \citep[V.5.1132b31-34]{Ethics}). According to  Joel Kaye, this means that Aristotle took a very different view of the purpose of economic exchange, it is performed  to correct for inequalities in endowment and to establish a social equilibrium and not in order to generate a profit. This view is at odds with that taken by mainstream modern economists, or even anthropologists such as Graeber, who seem to be committed to a nineteenth century ideal of ergodicity and determinism and so can ignore the centrality of reciprocity in markets.    This paper argues  that in the presence of uncertainty, society needs reciprocity in order to function.

Reciprocity and fairness are linked, and the importance of fairness in human societies is demonstrated  in the so--called `Ultimatum Game', an important anomaly for neo-classical economics \citep{T_AUG}.   The game involves two participants and a sum of money.  The first player  proposes how to share the money with the second participant.  The division is made only if the second participant accepts the split,  if the first player's proposal is rejected, neither participant receives anything.   The key result is that if the money is not split `fairly' (approximately equally)  then the second player rejects the offer.  This  contradicts the assumption that people are rational utility maximising agents, since if they were the second player would accept any positive payment.  Research has shown that  chimpanzees  are rational maximisers while   the willingness of the second player to accept an offer is dependent on age and culture.  Older  people from societies where exchange plays a significant role are more likely to demand a fairer split of the pot than young children or adults from isolated communities (\citep{MS_UBCA}, \citep{H_CPAHS}, \citep{CJM_CRM}). Fair exchange appears to be learnt behaviour developed in a social context and is fundamental to human society.

The relevance of the Ultimatum Game to the argument presented here is that separates the Classical themes of fairness and reciprocity from the Romantic theme of Darwinian `survivial of the fittest'.   In \emph{An Inquiry into the Nature and Causes of the Wealth of Nations}, published in 1776,   Adam Smith, working in the Classical framework, argues that humans are distinctive from other animals in the degree to which they are co-operative
\begin{quote}
Nobody ever saw a dog make a fair and deliberate exchange of one bone for another with another dog.\ent{WoN}{Book 1, Chapter 2}
\end{quote}
Humans , on the other hand, exhibit
\begin{quote}
 the propensity to truck, barter, and exchange one thing for another.\ent{WoN}{Book 1, Chapter 2}
\end{quote}
Markets are not simply a technical tool to facilitate life, but they capture a key distinction between humans and  other animals.  This is observation is very different to  Darwin, who wrote in  \emph{The Descent of Man} in 1871,  
\begin{quote}
 My object in this chapter is to shew that there is no fundamental difference between man and the higher mammals in their mental faculties.\ent{D_DMSRS}{p 36}
\end{quote}
Darwin would go on to note that 
\begin{quote}
 the weak members of civilised societies propagate their kind. No one  \ldots will doubt that this must be highly injurious to the race of man \citep[p 168]{D_DMSRS}
\end{quote}
and to argue that society should control this process within the ethic of Consequentialism.  Darwin did acknowledge that `survival of the fittest' did not explain the `nobler' aspects of civilisation \citep[p 161--167]{D_DMSRS}, but his arguments are integral to the  `social Darwinism' of Spencer and Galton that explained how the `best' became the ruling.  Darwinian metaphors are still a powerful feature of the paradigm centred on neo-classical economics and Consequentialist Ethics and the Ultimatum Game, and the concept of fairness, is an anomaly for the whole of this paradigm.

We conject that balanced reciprocity should be an axiom of exchange, possibly in preference to the axiom that the aggregate demand price is equal to the aggregate supply price \citep[Ch 2, VII]{K_GT}, or that a price is the cost of labour and a risk premium (e.g. Duns Scotus \citep[p 140]{K_ENFC}, Smith, Marx).  Reciprocity is anterior to the concept of justice,  in particular the argument for  justice and fairness as being essential components in commerce follow  Aristotle's arguments for how people, in a  state based on egalitarianism, can live together in an urban society.  On the basis of this axiom, we suggest, that the FTAP could be deduced, mirroring the analysis of Cardano,  Pascal-Fermat, Huygens.

The purpose of this section has been to give an explanation as to why the significance of fairness was lost in the pricing of contracts.  It argues that Western culture has oscillated between worrying about scarcity and uncertainty.  By comparing Appadurai and Graeber, we have concluded that when uncertainty dominates thinking,  reciprocity, and fairness, become important.  The final section considers the implications on financial economics of acknowledging the fundamental  role of uncertainty in economics and the consequential adoption of balanced reciprocity as an axiom.

\section{Implications for Financial Economics}

We can identify three immediate consequences of regarding the Fundamental Theorem of Asset Pricing as a consequence of the axiom of balanced reciprocity. 

The first consequence is related to the EMH.   Miyazaki observes \citep[p 404]{M_BAS} that speculation by arbitrageurs has been justified as ensuring that markets are efficient.  The EMH is based on the axiom that the market price is determined by the balance between supply and demand, if this axiom is replaced by the axiom of reciprocity, the justification for speculative activity in support of efficient markets, disappears.  In fact, the axiom of reciprocity would de-legitimise `true' arbitrage opportunities, as being unfair.  This would not necessarily make the activities of actual market  arbitrageurs illicit, since there are rarely strategies that are without the risk of a loss,  however, it would place more emphasis on the risks of speculation and inhibit the hubris that has been associated with the prelude to the FC.

The second consequence concerns the incompleteness of markets, related to the work of Arrow and Debreu.  An argument in favour of the explosion in traded assets that has occurred since 1971 has been  that more assets leads to `more' completeness of the market, and in a complete market all claims can be hedged, and so there is no uncertainty in the market.  The most obvious, mathematical, objection to this argument is that completeness either exists or it does not, and increasing the number of assets is not going to solve the basic problem, since in continuous time we need to assume that there are an infinite number of future states of the economy.  Recognising that uncertainty cannot be avoided, and so reciprocity and fairness are needed, rather than creating more and more traded assets, is an important implication of the argument in this paper.

Mathematics is responding to the problem of incompleteness by investigating so-called `robust methods', robust to epistemic uncertainty.  A highly incomplete summary of this work  includes \citep{ALP_PHDS}, \citep{L_UVRFS}, \citep{FK_OAMU}, \citep{CO_RPH}, \citep{NS_SHDRMVU}, \citep{H-LT_MMMM}.  The general conclusion of this research is that the prices obtained from robust methods would inhibit trading in many products, since the bid-ask spread would be so high as to remove any prospect of  profit  in trading.

The third issue is related to the mathematical understanding of pricing in incomplete markets.  Within mathematics, martingale measures can be constructed outside the process of hedging that was employed by Bronzin, Thorp and Kassouf or Black, Scholes and Merton. Rama Cont and Peter Tankov make the following comment
 \begin{quote}
 Unless the martingale measure is a by-product of a hedging approach, the price given by such martingale measures is not related to the cost of a hedging strategy therefore the meaning of such `prices' is not clear. \citep[10.5.2]{CT_FMJP}
\end{quote}
Considering the FTAP as a statement derived from reciprocity resolves this crux, which is difficult to interpret in the context of other price axioms.

The legitimacy of speculation and its relationship to gambling has been significant in the development of the markets since 1971, and needs to be considered.   In 1968 the Chicago Board of Trade  consulted lawyers about offering an index future, but had been told it would probably be ruled as illegal.  While grain, soybeans, shrimp, onions and even stocks in IBM could be delivered, the `index' could not, it was simply a number the Dow Jones published each day.  There was no way, in the lawyers' opinion, the Illinois courts would consider bets on an abstract number legitimate speculation\ent{ENC}{p 145}.  However since all the  parameters of the Black-Scholes equation were `known'  and the function was deterministic, the value of the option was known, up to the limits of epistemic uncertainty.  The removal of uncertainty in pricing options had a dramatic effect, it meant that options trading was not gambling, since there was, apparently,  no randomness in the process.  MacKenzie discussed its effect with the legal counsel to the Chicago Board Options Exchange at the time, Burton Rissman, who made the point that
\begin{quote}
Black-Scholes was what really enabled the exchange to thrive \ldots we were faced in the late 60s -- early 70s with the issue of gambling.  That fell away, and I think Black-Scholes made it fall away.  It wasn't speculation or gambling it was efficient pricing. \ldots I never hear the word ``gambling'' again in relation to stock options traded on the Chicago Board Options Exchange\ent{ENC}{p 158}
\end{quote}
More recently, the Potts Opinion on the nature of Credit Default Swaps, argues that a CDS is not a `wager' since the parties involved are not betting on the whether, or not, a firm will default on its bonds; if they do not hold the bonds, they are `speculating' in the Appadurai/Brenner \& Brenner sense.

At the heart of modern finance is `gambling', whether on an outcome or on a mis-pricing.  While Cardano investigated the act of gambling, concluding that its morality rested in the character of the participants; their willingness and knowledge, his analysis might  be considered weak today.    Gambling had been outlawed in the medieval period, often because it was felt that young men spent time gambling when they should have been participating in military training {\citep[p 58]{B_GS}}. The seventeenth century economist, Sir William Petty, had observed that lotteries were  ``a tax upon unfortunate, self-conceited fools'' and in 1699, lotteries were banned temporarily on the grounds that they had
\begin{quote}
 most unjustly and fraudulently got \ldots great sums of money from the {children} and {servants} of several gentlemen, traders and merchants \ldots to the utter ruin of many families \cite[p 11]{B_GS}.
\end{quote}  

Modern day prohibitions on gambling emerged alongside Consequentialism. From the start of the eighteenth century, views expressed by Petty evolved into what  Lorraine Daston has described as 
\begin{quote}
 Recurring motifs [of] the waste of time and money; the neglect of familial and business duties; the erosion of social trust;and the severed link between hard work, talent and gain.  But at the heart of the moral critique of gambling was an emotional portrait of the gambler as one racked by uncontrollable passions \ldots Neither rational self-interest nor conscience could any longer be depended upon to check wild impulses or restrain impetuous desires {\citep[p 161]{D_CPE}}
\end{quote}
 
 In 1774 the British Parliament passed the Life Assurance Act that prohibited individuals taking out  insurance policies unless they had an `interest' in the life of the insured, and this Act became known as the  `Gambling' Act.  Lotteries  were a necessary tool of public finance that prevented the stagnation and crises suffered by states reliant on taxation\ent{N_ESCE}{}. By the start of the nineteenth century, finance had developed to such an extent that governments could tax more effectively, notably the incomes of the middle classes (first introduced  in 1798), or to borrow from the middle and upper classes.  As a consequence,  the working classes could be excluded from the opportunities  to get rich that participating in public-finance, by purchasing lottery tickets, provided.   In 1808 the British Parliament set up a committee to ``inquire how the evils attending Lotteries have been remedied by the laws passed''.  The parliamentarians concluded that, despite the fact that the British government was still raising money through lotteries, ``the foundation of the lottery system \ldots  under no \ldots regulations \ldots will it be possible \ldots[to] divest it of \ldots evils'' {\citep[p 12]{B_GS}}.  The status of lotteries was changing and in 1823 they were outlawed, with the last draw taking place in 1826.

One explanation for the de-legitimisation of lotteries, and gambling in general, comes from  Brenner and  Brenner  who make the point that during the seventeenth and eighteenth centuries there was significant social and economic change.  In this environment gambling and speculation provided the `lower classes' with a means to to climb up the social ladder, to the horror of the ruling classes {\citep[pp 98--104]{B_WC}}.  In particular, chance flouted God by rewarding the undeserving, an issue in 2004 when there was a public outcry against a convicted rapist winning \pounds 7 million in the UK's National Lottery, and flouted science by disrupting Darwinian--Platonic arguments  of `survival of the fittest', which explained to the nineteenth century ruling classes why they ruled.  To justify the de-legitimisation of the lottery,  working-class gambling was linked to other vices, and from around the middle of the eighteenth century there was  an increase in moral criticism of gambling, which  persisted for the next two hundred years.   According to  Marieke de Goede,
\begin{quote}
 The evil of gambling was inserted into a web of meaning that linked gambling to other vices, including drunkenness, crime, and prostitution.  Lottery critics complained that the prolonged lottery draws were and side betting were distracting Londoners from their proper labours\ent{dG_VFF}{p 55}
\end{quote}

In the context of ergodic economics, the idea that financial markets are beneficial, let alone correct for inequalities seems perverse.  However anthropologists have long recognised the importance of gambling in archaic societies appearing in the Vedic scriptures; Greek mythology; potlach ceremonies; aboriginal Australia and New Guinea; the modern Hazda\ent{S_SAE}{p 27}.  To appreciate the role gambling plays in these societies,  consider a case observed in an Australian aboriginal group, the Momega in  Arnhem Land around 1980.  The community had access to social security payments and there was often a surplus left over after  essentials had been bought.  Jon Altman, studying the group, observed
\begin{quote}
 this surplus was not equally bestowed. \ldots This variability in bestowal was extremely arbitrary and it resulted in inter--household variability in access to cash.\ent{A_GA}{p 56}
\end{quote}
This variability can  seen as a subjective discrimination of the community by the Australian government. Gambling, according to Altman, ``acted effectively to both redistribute cash \ldots [it] provided a means for people with no cash income to gain cash''\ent{A_GA}{pp 60-61}, from a small stake a larger cash reserve could be generated.  The random distribution created by gambling, while not uniform, some would lose, some win, was none the less objective and most people ended up with `a fair share' of the cash. This was important in  non-hierarchical communities because it meant that one arbitrary bestowal of money was not corrected by another subjective distribution, such as redistribution by a chief.

Another anthropologist, William Mitchell, considered the role that gambling plays in disrupting hierarchical social structures, such as the Indian caste system, by studying the Wape in New Guinea around the same time
\begin{quote}
 An important task of Dumont's classic study
of Indian caste was to demonstrate how inequality is maintained. My task is the obverse, that
is, to reveal how the Wape defeat the formidable principle of hierarchy to maintain male equality.
How do the Wape, who, as individuals, desire wealth and who, since the 1930s, have been
directly tied to a world capitalist market system, prevent wealth from being successfully manipulated
by a few men to raise themselves above others? The paradoxical answer is deceptively
simple: through gambling.\ent{M_DH}{}
\end{quote}
This is an explanation for the pervasive nature of gambling in neolithic communities: it is an objective, fair, mechanism for the redistribution of wealth.  

This does not mean that gambling is benign.  De Moivre and Montmort studied `ruin probability' and `time to ruin' problems, these reveal that if there large differences in gamers wealth, the richer gambler will be able to ruin all other players and walk away considerably wealthier.  This means that we return to Cardano's original observation, that ``The most fundamental principle of all in gambling is simply equal conditions, e.g., of opponents \ldots''.

Modern mathematics captures a problem of combining gambling with Mill's advice that economics should be  concerned  solely with judging of the comparative efficacy of means for obtaining as much wealth as possible.  In  the aftermath of the FC, Hanqing Jin and Xunyu Zhou investigated the concept of `greed'  using mathematics \citep{JZ_GLPL}.  The paper defines greed as a ratio between a person's current wealth and a, higher, reference target level.  The mathematics told them that as the target level approached infinity, the probability of reaching that target did \emph{not} approach zero.  So, as the reference grew the person would become more `aggressive' to keep the  wealth target in sight,  and a rational agent maximising \emph{expected} utility would borrow more and more in order to  achieve impossible targets.  

While offering an obvious warning to Mill's programme, the work of Jin and Zhou  goes further, it  undermines the very belief in the  `calculability' of economic consequences, because their work was only made possible on the basis of their 2008 paper   \emph{Behavioral portfolio selection in continuous time} \citep{JZ_BPSCT} which came with the observation that
\begin{quote}
 a lack of study of  continuous--time behavioural portfolio selection is certainly not because the problem is uninteresting or unimportant; rather it is because, we believe that, the problem is \emph{massively} difficult as compared to the conventional expected utility maximisation problem
\end{quote}
Before 2008 the mathematics simply did not exist to analyse continuous--time portfolio selection using `S-shaped' utility functions, first proposed in 1948 by Friedman and Savage \citep{FS_UACIR}. The assumption that just because we are rational, we can calculate the correct answer pre-supposes that the mathematics to perform the calculation is in place.

Cardano, concluded that a just gamble involved ``willing and knowledgeable players'', the act and consequences of gambling were irrelevant, it was the character of the gambler that was important.   The journalist and social anthropologist, Gillian Tett, touches upon this in her book, \emph{Fool's Gold}, that discusses the role of the investment bank J. P. Morgan in the FC.   Tett explains that at J. P. Morgan
\begin{quote}
 the senior bankers still talked about banking as a noble craft, where long--term relationships and loyalty mattered, both in dealing with clients and inside the bank. \ldots The young trainees on the training programme were solemnly told that while the bank would tolerate `errors of judgement', an `error of principle' was a sacking offence.  `First class banking [in a first class way]' remained the mantra\ent{T_FG}{p 17--18}
\end{quote} 
The bank was involved in many of the technical developments that were subsequently held to account for the crisis: the risk-metrics Value at Risk; the use of the Gaussian copula to model dependencies between risks; and the Credit Default Swap.   Tett's narrative is that in  2005--2006  J. P. Morgan's shareholders were putting the bank's managers under intense pressure to mimic the revenues being reported by other investment banks, who were actively investing in CDOs of MBS.
\begin{quote}
[The J.P. Morgan chief executive] made it clear that he wanted a mortgage production line, so Winters had duly asked his staff to re--examine how to create a profitable business selling mortgage--based CDOs.

When they crunched the numbers, though, they ran into a problem.  ``There doesn't seem to be a way to make money on these structures,'' Brian Zeitlin, one of the bankers who worked in the CDO division reported. \ldots

Reluctantly, Winters told the J.P. Morgan  management should not open the spigots on its pipeline after all.  \ent{T_FG}{pp 148--151}
\end{quote}
The explanation is that J. P. Morgan, in developing the tools associated with the FC, had embedded skills in the organisation that enabled it to understand the risks.  There seems to be a relationship between the ethos of ``First class banking [in a first class way]'' and using science to guide speculation, successfully.

What this account does not deal with is the problem of the `willingness' of the participants, particularly in the case of an asymmetry between counterparties.  This problem is at the heart of Aquinas' original analysis of the merchant and the starving citizens, there is an imbalance in the status of the merchant, with knowledge, and the citizens, without it.  This imbalance between parties was understood by Aristotle, and expressed in the ``equal conditions'' of Cardano, and means that balanced reciprocity may not be appropriate.  This issue,  `inequality of bargaining power', has been incorporated into English and U.S. law (\emph{Lloyds Bank Ltd v Bundy}, [1974] EWCA Civ 8, \emph{Williams v. Walker-Thomas Furniture Co.}, 350 F.2d 445 (C.A. D.C. 1965)) and can be accommodated within Virtue Ethics, as will become clear in what follows.

Cardano came to his conclusions within the context of medieval Catholic Virtue Ethics:  the four `Pagan' or `Cardinal' virtues; Courage (\emph{Fortitudo}); Justice (\emph{Iustitia}); Temperance (\emph{Temperantia}); and Prudence(\emph{Prudentia}), which originated in \emph{Nicomachean Ethics}, and  three `Christian' virtues; Hope (\emph{Spes}); Faith (\emph{Fides});  and Charity (\emph{Caritas}).  Medieval scholars approached morality using the same framework that they used to study physics or  medicine by  blending  four elements, or humours, in the right manner.  For example, Charity and Faith yield loyalty, Temperance and Courage give humility and Justice, Courage and Faith result in honesty\ent{M_BV}{p 361}. An ethical life was one that exhibited all, not just some, of the virtues, and a merchant, by demonstrating them,  could be seen as being as virtuous as  a prince or a priest.  This was not just a Latin Christian view, the first century Mahayana Buddhist \emph{Vimalakirti Sutra} tells the story of how a virtuous merchant teaches kings and monks.

Following the analysis of \emph{Nicomachean Ethics}  medieval scholars, like Aquinas and Olivi, placed the virtue of Justice at the centre of commerce.  Prudence is the common sense  to decide between different courses of action, it is at the root of reason and rationality and can be seen as the motivation for all science.  This virtue is the one most closely associated, in the modern mind at least, with effective merchants.  Temperance, a word that comes into English with Grosseteste's translation of \emph{Ethics}, is the virtue least associated with modern bankers.  However, the modern understanding of temperance as denial or abstinence is not only how a medieval friar would have understood the virtue. The word is related to `temper' and is concerned with getting the right balance between the virtues. A good merchant would exhibit the virtue by mixing Courage, to take a risk, and Prudence, allowing for the unforeseen, and diversifying.  

Faith is the ability to believe without seeing, and was central to Olivi's whole philosophy.  The Latin root is \emph{fides}  captures the concept of trust, the very essence of finance exhibited by the ``promise to pay''. While Faith is backward looking, you build trust, Hope is its forward-looking complement.  Charity, along with Temperance, is the virtue least likely to be associated with merchants.  While we now think of  charity in terms of giving to others, in the past it was associated with a love, or care, for others.     Shakespeare's play \emph{The Merchant of Venice} is about  `Antonio, a merchant of Venice' who characterises Charity, or \emph{agape}, though his sacrifices for his young friend Bassanio.  The view that Antonio and Bassanio were physical lovers is a modern interpretation that does not distinguish  \emph{storge} (familial love, the deficit Jessica/Shylock), \emph{philia} (friendship, Portia/Nerissa, Lorenzo/Bassanio), \emph{eros} (physical love, Portia/Bassanio, Lorenzo/Jessica) and \emph{agape} (spiritual love, Antonio/Bassanio, Shylock's deficit), clear themes running through the play.  We would suggest that the problem Graeber should be tackling in his discussion of debt is not the presence of reciprocity but rather the absence of Charity.  

This point could have  had a role in both the FC and in the failure of Long Term Capital Management (LTCM) in 1998.  If a lender has care for the borrower, they will not lend beyond the borrower's means.  The mechanisation of the credit-debit relationship in the lead up to the FC, through computerised credit scoring and securitisation, a combination of Deontological and Consequentialist morals, had the effect that the debtor was distanced from the creditor and enabled the sub-prime borrower to justify default and the evaporation of reciprocity.  With regard to LTCM, MacKenzie reports that on 2 September the firm informed investors that they had suffered losses, but the market was so dis-located that it was full of opportunities,  and so requested more funds.  LTCM's counterparties took a different view
\begin{quote}
When it became apparent they [LTCM] were having difficulties, we thought that if they are going to default, we're going to be short a hell of a lot of volatility. So we'd rather be short at 40 [at an implied volatility of 40 per cent per annum] than 30, right? So it was clearly in our interest to mark at as high a volatility as possible. That's why everybody pushed the volatility against them, which contributed to their demise in the end.\ent{M_LTCMSA}{p 366}
\end{quote} 
In this case, LTCM's counter-parties seem to be ``asking for more than what is due'', and the principle of reciprocity on which the financial system is constructed, is lost, resulting in a near collapse of the system.

In summary, Charity will address the problem of `inequality of bargaining power'.

Theologians often remark that faith is not the opposite of doubt, but its complement; if the existence of God is indubitable, faith is redundant.  Similarly, Hope and Courage are unnecessary if we know what will happen, though Prudence still has a role; if a warrior knows that they will be killed in a fight, it is prudent to avoid the battle.  While Courage and Prudence, pagan virtues, and Faith and Hope, Christian virtues, directly address uncertainty, Justice and Charity tackle risk by creating the social structures that enable diversification and risk sharing that is exhibited in reciprocity.  Temperance is about establishing the right mix of Courage and Prudence, Faith and Hope and Justice and Charity.  The seven virtues can be seen as directly addressing uncertainty.

%Commercial activity was not alien to Scholastic scholars and one of the earliest descriptions of an entrepreneur comes in `On Contracts and Usury' written between 1431 and 1433 by San Bernardino of Siena.  Bernardino was an ascetic Franciscan who seems to have been familiar with Olivi's writing  (\citep[p 81]{ETBAS}, \citep[p 118]{K_ENFC}) and would be one of the inspirations behind Savonarola's Christian Republic in Florence and its Bonfires of the Vanities. Bernardino, like Olivi, recognised the important service that merchants provide in supplying society with the goods it needs, when and where they are required.  He realised that to be successful a merchant had to be well informed, of prices, the qualities of goods and the market; be diligent in keeping accounts; be hard working; and, importantly, be willing to take on risk.  Bernardino observed that there were very few people who had all the these qualities that were needed to be a profitable merchant \citep[pp 81--82]{ETBAS}.  

%and, as such, it would be at odds with  Appadurai'sassessment that the  modern speculator is \begin{quote} avaricious, adventurous, exuberant, possessed, charismatic, excessive, or reckless  \citep[p 536]{A_GFM}\end{quote}

\singlespacing

%\bibliography{all}

\end{document}